\pdfoutput=1
\documentclass[showpacs,twocolumn,prl]{revtex4}

\usepackage{epsfig}
\usepackage{amsmath}
\usepackage{graphicx}
\usepackage{amssymb} 
\usepackage{graphics}

\begin{document}

\title{Channel-Width Dependent Enhancement in Nanoscale Field Effect Transistor}

\author{Xihua Wang, Yu Chen, Mi K. Hong, Shyamsunder Erramilli, Pritiraj Mohanty}

\affiliation{
Department of Physics, Boston University, 590 Commonwealth Avenue, Boston, MA 02215}


\begin{abstract}
We report the observation of channel-width dependent enhancement in nanoscale field effect transistors containing lithographically-patterned  silicon nanowires as the conduction channel. These devices behave as conventional metal-oxide-semiconductor field-effect transistors in reverse source drain bias. Reduction of nanowire width below 200 nm leads to dramatic change in the threshold voltage. Due to increased surface-to-volume ratio, these devices show higher transconductance per unit width at smaller width. Our devices with nanoscale channel width demonstrate extreme sensitivity to surface field profile, and therefore can be used as logic elements in computation and as ultrasensitive sensors of surface-charge in chemical and biological species.
   
\end{abstract}


\maketitle

The foundation of semiconductor microelectronics industry has relied on superior material properties of silicon and its scaling properties. Combination of silicon with the field-effect transistor (FET) design has allowed devices that are smaller, faster and cheaper. However, the traditional scaling of conventional planar CMOS devices leads to performance limitation due to increased leakage current with reduced gate thickness \cite{itrs01}. There has been intense activity in both silicon \cite{Arden03} and non-silicon based  materials \cite{CNT-logic,CNT1,CNT2} and gate dielectrics for improved performance; however, investigation of alternative device architecture with silicon remains a major thrust, due to large-scale manufacturing constraints \cite{Arden03}.   

In comparison to bulk-planar transistor configuration, vertical three-dimensional channels between the source and the drain can provide better control of current flow within the channel, better $I-V$ characteristics and better sub-threshold slopes in specific bias configurations. In addition to these improved properties, control of transverse channel-width dimension rather than the gate thickness offers a new method of improving performance. We show that in a vertical 3D transistor, the performance can be enhanced as the channel width is scaled down. Thus we avoid the problem of increased leakage current seen in conventional devices where the gate thickness is reduced. 

Vertical three-dimensional channel structures are naturally obtained in carbon nanotubes (CNT) or chemically-grown silicon nanowires (SiNW) mounted between source and drain electrodes. Bottom-gated FET devices using CNT and SiNW have been used to demonstrate logic operation as well as molecular sensing \cite{Lieber01, Lieber04, Gruner03}. Apart from the issue of lack of materials control, the main difficulty with this bottom-up approach continues to be the fundamental lack of control in the critical dimension (CD) that determines the device sensitivity and response. 

In this letter, we report fabrication and measurement of $I-V$ characteristics of novel field effect transistors with three-dimensional silicon nanowire channels between the source and the drain. Starting from Silicon-on-Insulator (SOI) wafers, we use a top-down approach for fabricating nanomachined three-dimensional silicon nanowires with controlled channel width, using electron-beam lithography \cite{Mohanty06, Reed07, Linnros07, Vogel04}. The CMOS-compatible nanofabrication approach is suitable for scalable manufacturing because the critical dimension of the channel width as well as the gate configurations are physically engineered. When the device is operated in reverse source drain bias, the threshold voltage demonstrates dramatic change as the channel width is reduced below 200 nm. The device performance, evaluated in terms of transconductance per unit width of nanowire surfaces, is enhanced as the channel width is reduced. This nanoscale-dependent behavior can be exploited in device applications as logic elements and sensors \cite{Mohanty08, yu-biosensor}. 

The nanowire channel in our field effect transistor devices are fabricated using $e$-beam lithography followed by surface nanomachining. The starting SOI wafer consists of 100 nm thick silicon as device layer, 380 nm silicon dioxide insulation layer, and 600 $\mu$m of silicon substrate. The \textit{p}-type device layer volume resistivity is $10-20 \mbox{ $\Omega$$\cdot$cm}$, corresponding to $~10^{-15}$ cm$^{-3}$ initial doping. The depletion width for bulk silicon (Si) with the same doping concentration is $L_D=130 \mbox{ nm}$. The device layer of SOI is fully depleted, since its thickness $H=100 \mbox{ nm}<2L_D$. There is no further intentional doping or high-temperature annealing. The advantages of not performing this step are that the fabrication process is easy and we do not have to worry about the nonuniform doping profile in nanowires due to source/drain doping or high-temperature annealing process. The drawback is that it leads to non-Ohmic contact, which may reduce the device performance. However, we have developed a simply analytical model to extract all parameters for evaluating device performance. We also find that the equivalent resistance of the non-Ohmic contact is much smaller than the nanowire resistance when small reverse bias is applied, so the device performance is barely influenced by the non-Ohmic contact in the sensing applications.      

\begin{figure} []
	\includegraphics[scale=0.5]{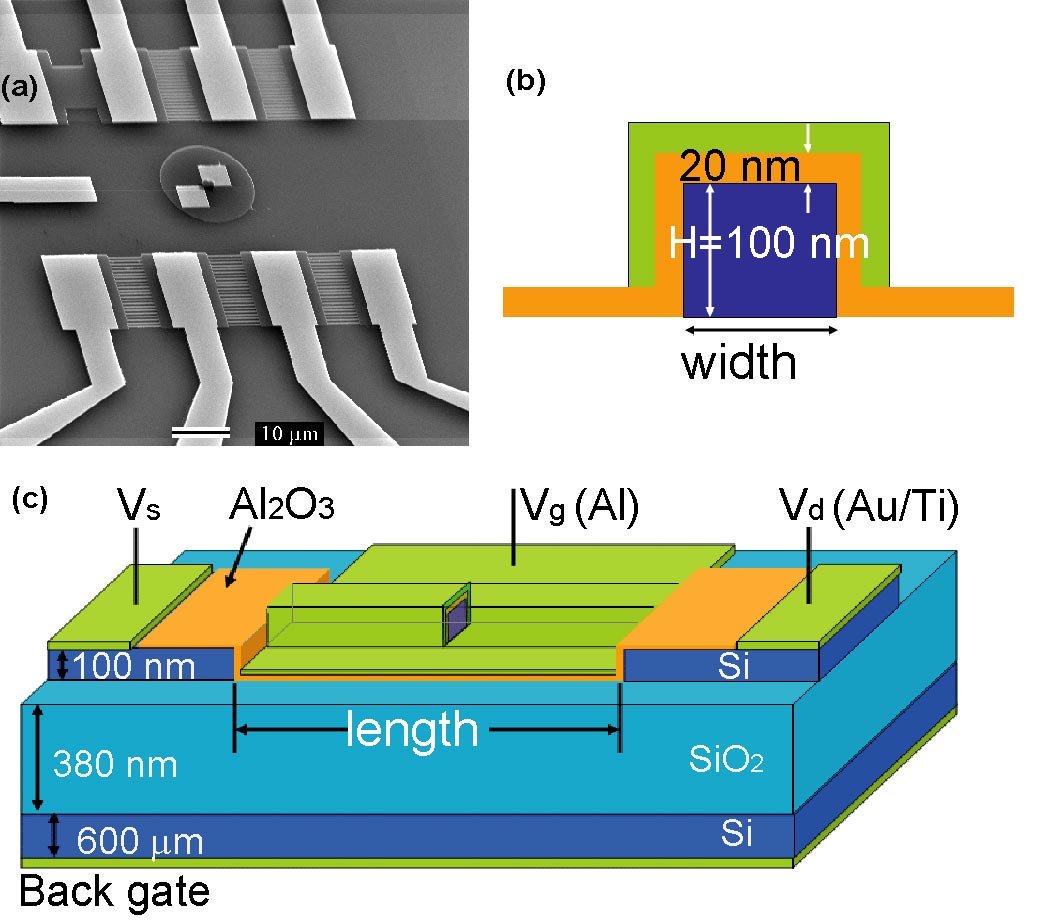}
	
	\caption{ (a) Scanning electron micrograph of the device before depositing the top electrode. (b) Schematic cross-sectional view of the nanowire showing the top and side surfaces, totally three surfaces, are contacted with the top gate. (c) Perspective, showing the source, drain and gate electrodes. Schematic representations are not drawn to scale.  }
	
	\label{fig1}
\end{figure}

Fig.~\ref{fig1} (a) shows a scanning electron micrograph of the devices. Fig.~\ref{fig1} (b)-(c) shows schematic views of the nanowire. The nanowire channel dimension are length $L$, width $w$, and thickness $H = 100$ nm. A 20-nm thin dielectric film of Al$_2$O$_3$, grown by atomic layer deposition (ALD), serves as an insulation layer between the top gate and the silicon nanowire. The three sides of the device are all covered with aluminum electrode as the top gate, designed to provide gate control of the channel current. The top gate voltage $V_g$ is controlled while the back gate is either floated or grounded. 

\begin{figure} [t]
	\includegraphics[scale=0.45]{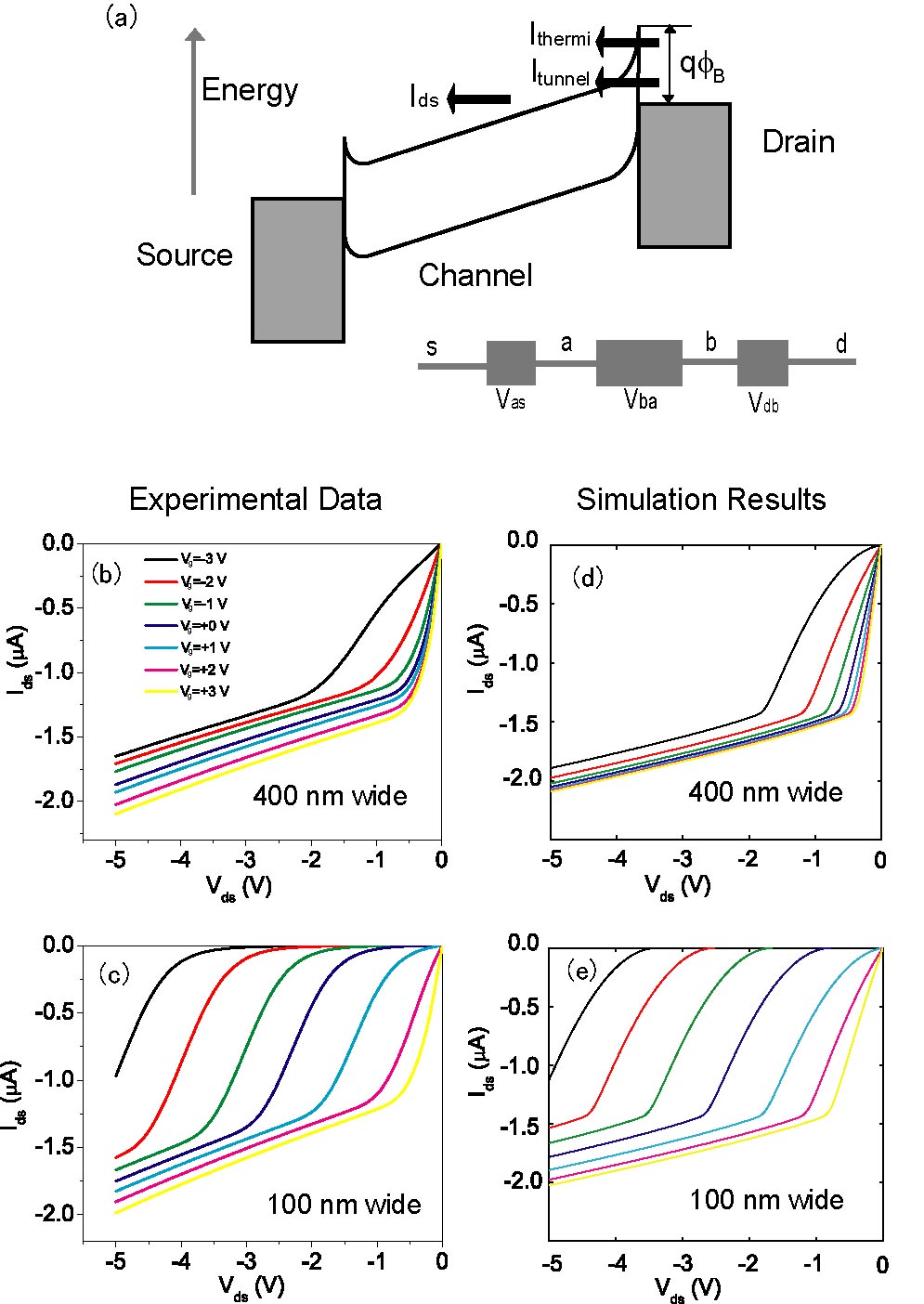}
	
	\caption{(a)  Band diagram showing the direction of current flow. The contact potential drops are in series with the intrinsic nanowire potential drop.  (b) (c) show I$_{ds}$ vs. V$_{ds}$ measurement for silicon wires with the width 400 nm, and 100 nm, respectively. The back gate is floated in the measurement. (d) (e) show simulation results at negative bias I$_{ds}$ vs. V$_{ds}$ for silicon wires with the width 400 nm and 100 nm, respectively.}
	
	\label{fig2}
\end{figure}

Fig.~\ref{fig2} (a) displays the band diagram for reverse source drain bias situation. The experimentally measured source-drain voltage $V_{ds}$ is a series combination of the contact potential drops and the intrinsic voltage drop $V_{ba}$ across the nanowire. 

Fig.~\ref{fig2} (b) (c) show the $I-V$ characteristics of silicon nanowire FET at different widths when the  back gate is floated. The general trend can be appreciated by considering the $I-V$ curves for the 100 nm wide device, when the top gate voltage $V_g$ is held at zero (dark blue curve). As $V_{ds}$ is decreased from 0 V, the current remains small until it reaches about -1V. As $V_{ds}$ is decreased further, the magnitude of $I_{ds}$  increases sharply. Below a $V_{ds}$ value of around $-3$V, the $I-V$ curves are linear, with a small constant slope. No leakage current is observed between gate and source/drain in the measurement. Fig.~\ref{fig2} (d) (e) demonstrate the simulation results for devices using the following analytic model. 

One Schottky barrier is formed between the source contact electrode and the silicon nanowire (SiNW) channel. Another barrier is established between the drain contact electrode and the SiNW channel. Silicon-based Schottky barrier FETs with large channel widths \cite{Larrieu04} have been studied extensively. Koo et al \cite{Vogel05} used a partially covered top gate to study the effect of  gate coupling under positive source drain bias. Here we study a fully covered top gate under reverse bias conditions, all parameters for evaluating device performance are extracted using the analytic model.

For narrow width wires, when the $V_g=0$ V, the conducting channel is assumed to be blocked when $V_{ds}=0$ V. The response of $I_{ds}$ to the gate voltage is characterized by a threshold voltage $V_t$, such that the channel is opened when  $V_g  > V_t+V_{ds}$.  In our model, the threshold voltage $V_t$ depends on the surface-to-volume ratio of the nanowire. The device can be seen as a forward-biased and a reverse-biased Schottky barrier in series with the silicon nanowire (see Fig. ~\ref{fig2} (a)). The experimentally measured voltage drop $V_{ds}$ is related to the intrinsic nanowire potential drop $V_{ba}$, and the contact potential drops $V_{db}$, $V_{as}$ at the drain and source electrodes respectively by $V_{ds}=V_{db}+V_{ba}+V_{as}$. When the device is operated in reverse source drain bias, $V_{as}$ can be neglected, and the equation is further reduced to $V_{ds}=V_{db}+V_{ba}$.

The current passing through the silicon nanowire channel is related to the intrinsic voltage drop $V_{ba}$ across it as
\begin{equation}
	I_{ds}=K[(V_{g}-V_{t})V_{ba}-\frac{m}{2}V_{ba}^{2}],
\end{equation}
where the gain parameter $K$ is $K=\mu C_{ox} \frac{W_{eff}}{L}$. Here $\mu$, $C_{ox}$ and $\frac{W_{eff}}{L}$ represent mobility, gate oxide capacitance per unit area, and effective width of nanowire surfaces over channel length, respectively. The correction factor $m$ allows us to take the body effect into account \cite{BodyEffect}. The current passing through the Schottky barrier can be modeled by a combination of the thermionic emission current $I_{thermi}$ and the tunneling current $I_{tunnel}$:
\begin{equation}
	I_{ds}=I_{thermi}+I_{tunnel}.
\end{equation}
The thermionic emission current is given by 
\begin{eqnarray}
\nonumber
I_{thermi}=AA^*T^2\mbox{ exp} \left(-\frac{e\phi_{B0}}{k_{B}T}\right)\left[ \mbox{ exp} \left(\frac{eV_{db}}{k_{B}T}\right)-1\right] \\ 
=C_1 \left[\mbox{ exp} \left(\frac{eV_{db}}{k_{B}T}\right)-1\right],
\end{eqnarray}
where $C_1$ is called the reverse saturation current parameter, $A$ is the contact area, $A^*$ is Richardson constant, $T$ is the temperature, $k_B$ is Boltzmann constant and $\phi _{B0}$ is the Schottky barrier height in the absence of an applied voltage.

The tunneling current can arise from two different contributions: a tunneling current across the Schottky barrier, or due to trap-assisted tunneling. In order to evaluate the different contributions, we first estimate the tunneling current across the Schottky barrier \cite{Park03,Asada99} using the Fowler-Nordheim tunneling expression for the associated current density $J_{FN}$:
\begin{equation}
	J_{FN}=\frac{e^2 (m_e/m_n^*)}{8\pi h \phi _b} E_b^2\exp\left(-\frac{8\pi \sqrt{2m_n^*(e\phi _b)^3}}{3ehE_b}\right),
\end{equation}
where $\phi _b=\phi _{B0}-\sqrt{\frac{eE_b}{4\pi \epsilon}}$. Here $m_e$ is electron mass, $m_n^*$ is the effective mass of electron, $h$ is the Plank constant, $E_b$ is the electric field in the Schottky barrier and $\epsilon$ is the permittivity. 
The Schottky barrier width can be estimated with full depletion approximation:
\begin{equation}
	x_d=\sqrt{\frac{2\epsilon (\phi _{B0}-V_{db})}{eN_p}},
\end{equation}
where $N_p$ is doping concentration. So $E_b=-V_{db}/x_d$. For electron conduction under reverse bias, we can use $\phi _{B0}=0.51$ V \cite{Taubenblatt84} for estimates. With sample-specific numerical values for the parameters, we obtain $\phi _B\approx \phi _{B0}$, and $J_{FN}<10^{-40}$ A/m$^2$. Therefore we can neglect the tunneling current. Considering the second possibility for tunnelling current contribution, due to Frenkel-Poole emission \cite{Yu06} which arises from trap-assisted tunneling, we write the corresponding current density $J_{FP}$ as: 
\begin{equation}
	J_{FP}=C_{FP}E_b \mbox{ }\texttt{exp}\left[-\frac{e(\phi _t-\sqrt{eE_b/\pi \epsilon })}{k_BT}\right].
\end{equation}
where $\phi _t$ is the barrier height for electron emission from the trap state. We combine the contributions from thermionic emission and Frenkel-Poole emission to model experimental data. The simulations are used to extract the reverse saturation current parameter $C_1=1.4$ $\mu$A. Based on our device contact area $A=100$ $\mu$m$\times$10 $\mu$m, room temperature, $T=300$ K, Richardson constant for silicon, $A^*=120$ A/cm$^2$/K$^2$, we get the Schottky barrier height between Ti and Si, $\phi_{B0}=0.47$ eV. This is in good agreement with the barrier height(0.51V)\cite{Taubenblatt84} for previous estimation. Further high temperature studies ($100-200\mbox{ }^oC$) show that both thermionic emission current and tunneling current increase as temperature increases. This confirms the tunneling current is from trap-assisted tunnelling.

\begin{figure} [t]
	\includegraphics[scale=0.45]{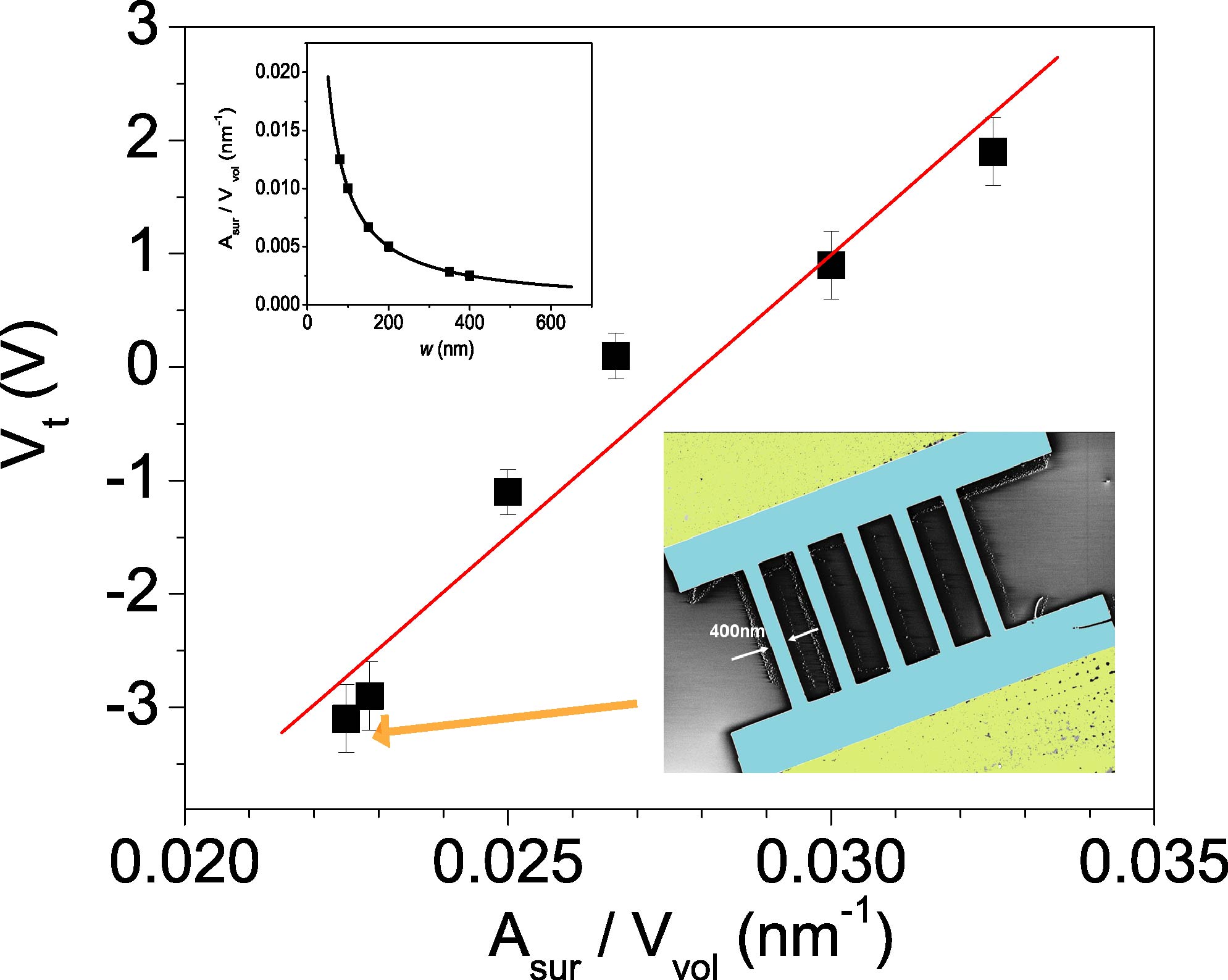}
	\caption{Threshold voltage as a function of surface-to-volume ratio for silicon wires. The solid line is a fit to experimental data (dots). The top inset shows surface-to-volume ratio for different widths, according to Eq.~\ref{ThSV}. The bottom inset displays a scanning electron micrograph of a device containing 400 nm wide silicon wires array.}
	\label{fig3}
\end{figure}

Using the analytic model, we can get values of intrinsic nanowire potential drop $V_{ba}$, and the contact potential drop $V_{db}$. We find that $V_{db}$ is much smaller than $V_{ba}$ as long as the absolute value of $V_{ds}$ is not big enough to reach the small constant slope region in the $I-V$ curve. So device performance is barely influenced by the non-Ohmic contact in the sensing applications. 

The threshold voltage derived from the experiment data is a crucial parameter for evaluating the width dependent effect, observed in the data. Fig. ~\ref{fig3} shows that as the nanowire width decreases, the threshold voltage increases. We parametrize the width dependence in terms of the effective surface-to-volume ratio, which is a function of the nanowire channel thickness, $H=100 \mbox{ nm}$, and width, $w$, with the relation 
\begin{equation}
	A_{sur}/V_{vol}=\frac{w+2H}{wH}=\frac{2}{H}\times \left(\frac{1}{2}+\frac{H}{w}\right).
	\label{ThSV}
\end{equation}
From Fig.~\ref{fig3}, we find a linear relation between the threshold voltages $V_{t}$ and the surface-to-volume ratio $A_{sur}/V_{vol}$ of silicon nanowires. 
\begin{figure} [t]
	\includegraphics[scale=0.7]{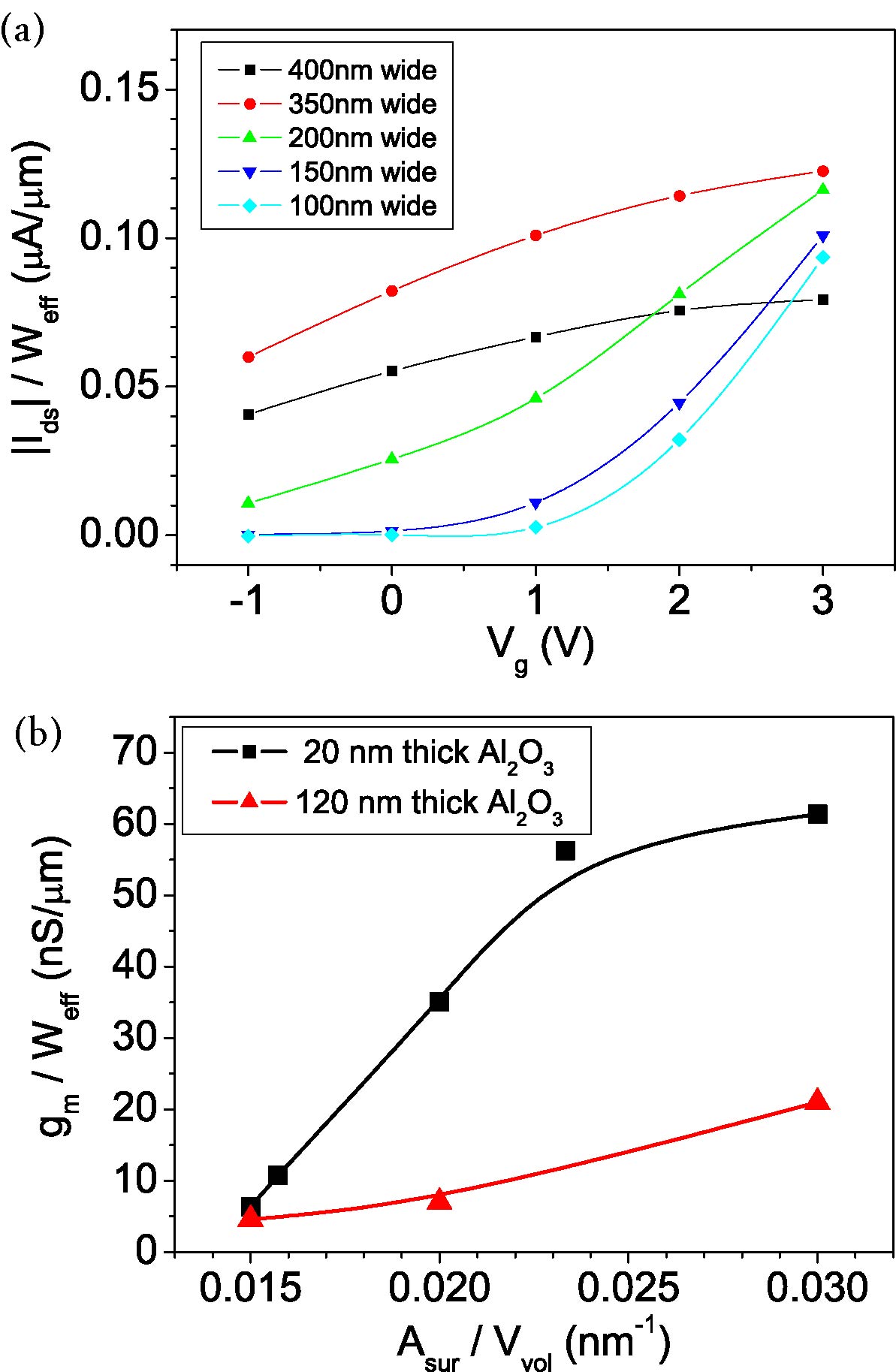}
	\caption{(a) Transfer characterisc plot ($V_{ds}=-0.1$ V) of the device with 20 nm insulation layer (Al$_2$O$_3$) thickness for different nanowire width. (b) Transconductance per unit width of the device as the function of surface-to-volume ratio.}
	\label{fig4}
\end{figure}

The most important parameter of evaluating device performance is the transconductance $g_m$. It is expressed as 
\begin{equation}
	g_m=\frac{\partial I_{ds}}{\partial V_g}=\mu C_{ox}\frac{W_{eff}}{L}V_{ds},
\end{equation}
in the linear region of transfer characteristic plot. We consider the electron inversion layer exists on the surface of nanowires. Thus $W_{eff}=(w+2H)\times n$, and $n$ is the number of nanowires in parallel between source and drain. Since diffferent nanowires have different $W_{eff}$, it is better to evaluate device performance using transconductance per unit width of nanowire surfaces. Fig.~\ref{fig4} (a) gives the transfer characteristics of nanowires with different width at $V_{ds}=-0.1$ V. The slop of the curve in the linear region is the transconductance per unit width for the device. These values of slops are plotted in Fig.~\ref{fig4} (b); the transconductance per unit width increases one order of magnitude (6.31 nS/$\mu$m to 61.4 nS/$\mu$m) as width shrunks from 400 nm to 100 nm. So the device performance is enhanced due to higher surface-to-volume ratio.  

The gate dielectric thickness (aluminum oxide layer) is important in conventional FET configurations. In our configuration, this thickness also plays a key role in determining the device response and sensitivity. Not only does the thicker oxide reduce $C_{ox}$, it also influences the geometry of 3D structure of nanowires. In Fig.~\ref{fig4} (b), we also plot transconductance of the devices with thicker oxide thickness (120nm). The competition between the two control variables, the gate dielectric thickness and the three-dimensional nanowire width, is clearly seen in the data. As expected, sensitivity decreases as the thickness increases, since the surface effect is suppressed as the oxide thickness becomes comparable to the transverse dimension of the wire. With the thick dielectric layer, devices with narrow wires still show reasonable sensitivity while devices with wider wires do not. 

Our approach combines two effects. The first is the expected enhancement in the control when the oxide layer thickness at the top gate is reduced. The second is a novel control of the channel flow by physically engineering a three-dimensional nanowire channel so as to increase the surface contribution. Both effects can be exploited for increased sensitivity and better device performance. In biosensing applications, the biomolecules binding to the modified nanowire surfaces cause a surface potential change, and this shift of surface potential behaves like changing top gate voltage. We have shown ultrasensitive protein detection and amplification of biomolecular recognition signal \cite{yu-biosensor}, as well as glucose sensing \cite{Mohanty08} with our novel transistor configuration. 

In conclusion, we have measured $I-V$ characteristics for silicon nanowire FET with different nanowire widths. The threshold voltage of the FET devices shows nanowire channel-width dependent effect. The device performance is enhanced due to increased surface-to-volume ratio. An analytical model by considering the Schottky barrier of the device provides a good explanation of the data, and is useful for evaluating device performance. Both experiment and the analytical model suggest that device performance can be enhanced by narrow channel width in a three-dimensional channel and thin dielectric layer. Our approach can be used for developing better transistors for logic operations and sensor applications in nanoelectronic semiconductor industry.


\end{document}